\documentstyle[prd,preprint,tighten,aps,epsfig]{revtex}
\newcommand{\beq}{\begin{equation}}
\newcommand{\eeq}{\end{equation}}
\newcommand{\bea}{\begin{eqnarray}}
\newcommand{\eea}{\end{eqnarray}}

\newcommand{\lsim}   {\mathrel{\mathop{\kern 0pt \rlap
  {\raise.2ex\hbox{$<$}}}
  \lower.9ex\hbox{\kern-.190em $\sim$}}}
\newcommand{\gsim}   {\mathrel{\mathop{\kern 0pt \rlap
  {\raise.2ex\hbox{$>$}}}
  \lower.9ex\hbox{\kern-.190em $\sim$}}}

\begin{document}
\draft
\preprint{
\begin{tabular}{r}
SFB 375-309\\
TUM-HEP 338/98\\
DFTT 71/98
\end{tabular}
}
\title{\bf Strange form factors of the proton: a new analysis
 of the $\nu$ ($\overline{\nu}$) data of the BNL--734 experiment}
\author{
W.M. Alberico$^{\mathrm{a}}$,
M.B. Barbaro$^{\mathrm{a}}$,
S.M. Bilenky$^{\mathrm{b,c}}$,
J.A. Caballero$^{\mathrm{d},\mathrm{e}}$,\\
C. Giunti$^{\mathrm{a}}$,
C. Maieron$^{\mathrm{f}}$,
E. Moya de Guerra$^{\mathrm{e}}$ and
J.M. Ud\'{\i}as$^{\mathrm{g},\mathrm{e}}$,
\vspace*{0.3cm}\\
\begin{tabular}{c}
{\it $^{\mathrm{a}}$INFN, Sezione di Torino  
and Dipartimento di Fisica Teorica,}\\
{\it  Universit\`a di Torino, 
Via P. Giuria 1, 10125 Torino, Italy}
\\
{\it $^{\mathrm{b}}$Joint Institute for Nuclear Research, Dubna, 
Russia}\\
{\it $^{\mathrm{c}}$Institut f\"ur Theoretische Physik, Technische 
Universit\"at Munchen,}\\
{\it D--85748 Garching, Germany }
\\
{\it $^{\mathrm{d}}$Dpto. de F\'{\i}sica At\'omica, 
Molecular y Nuclear,} \\
{\it Universidad de Sevilla, Apdo. 1065, E-41080 Sevilla, Spain}
\\
{\it $^{\mathrm{e}}$Instituto de Estructura de la Materia, CSIC,}
\\
{\it Serrano 123, E-28006 Madrid, Spain}
\\
{\it $^{\mathrm{f}}$Center for Theoretical Physics,}
\\
{\it Laboratory for Nuclear Science and Department of Physics,}
\\
{\it Massachussets Institute of Technology, 
Cambridge, MA 02139, USA}
\\
{\it $^{\mathrm{g}}$Dpto. de F\'{\i}sica At\'omica, 
Molecular y Nuclear,}
\\ 
{\it Universidad Complutense de Madrid, E-28040 Madrid, Spain}
\end{tabular}
}

\date{\today}
\maketitle
\begin{abstract}
We consider ratios of  elastic $\nu ({\bar\nu})$--proton cross
sections measured by the  Brookhaven BNL--734 experiment
and use them to obtain the neutral current (NC) 
over charged current (CC) neutrino--antineutrino asymmetry. 
We discuss the sensitivity of these ratios and of the asymmetry to
the electric, magnetic and axial strange form factors of the nucleon 
and to the axial cutoff mass $M_A$. 
We show that the effects of the nuclear structure and interactions on
the  asymmetry and, in general, on ratios of cross sections 
are negligible.
We find some restrictions on the possible values of the parameters
characterizing the strange form factors.
We show that a precise measurement of the neutrino--antineutrino 
asymmetry would allow the extraction of the axial and vector magnetic 
strange form factors in a model independent way.
The neutrino--antineutrino asymmetry turns out to 
be almost independent on the
electric strange form factor and on the axial cutoff mass. 

\end{abstract}
\vspace{0.5cm}
{PACS: 12.15.mn, 25.30.Pt, 13.60.Hb, 14.20.Dh, 14.65.Bt}

\newpage
\section{Introduction}

After the measurements of the polarized structure function of the proton,
$g_1$, in deep inelastic scattering\cite{Adams94,Abe95}, it  
 turned out, rather surprisingly, that the constant $g^s_A$, that 
characterizes the one--nucleon matrix element of the axial strange 
current, is of magnitude comparable  with the corresponding  $g^u_A$ and  
$g^d_A$  axial constants. A theoretical analysis of deep inelastic
data\cite{Ellis95} led to the following values for the axial  
constants: $g^s_A = -0.10\pm 0.03$,  $g^d_A =-0.43\pm 0.03$, 
$g^u_A= 0.83\pm 0.03$. In a more recent analysis of the
data\cite{Magnon}, the value $g_A^s=-0.13\pm 0.03$ was reported.
Though it is subject to several assumptions (extrapolation of $g_1$ 
to the $x=0$ point, SU(3)$_f$ symmetry, etc.), the rather large
value of $g^s_A$ stimulated new experiments on the measurement of deep
inelastic scattering of polarized leptons on polarized 
nucleons\cite{Adams97,Abe95b} and a lot of theoretical work
on the subject (see for example ref.\cite{Reya98}). 

Alternative approaches which allow  to obtain
the contribution of the strange quark current to the structure of the
nucleon were also developed\cite{Manohar,Alb96,PVelec}. 
Among these,
information on strange form factors of the nucleon can be obtained from
NC  scattering of $\nu(\bar\nu)$ on nucleons and 
nuclei\cite{Ahrens87,Garv93,Hor93,Barb96}. Up to now 
the most detailed investigation of NC  $\nu(\bar\nu)$ - proton scattering
 was done in the Brookhaven BNL--734 experiment.
>From the analysis of the data of this experiment 
a nonzero value of  $g^s_A$ was found\cite{Ahrens87}. 
This result, however, strongly depends on the value of the axial cutoff mass
$M_A$. For example, in the paper by Garvey {\it et al.}\cite{Garv93},
from the fit of the BNL data it was found:
$g_A^s = -0.21\pm 0.10$ and $M_A= 1.032 \pm 0.036$ GeV.
This fit, however, shows up strong correlations between the values
of $g_A^s$ and $M_A$: the data indeed are also compatible with $g_A^s=0$, 
 provided 
one assumes a slightly larger axial cutoff mass $M_A=1.086 \pm 0.015$~GeV,
which  is in agreement with quasielastic neutrino-nucleon data. 
It is clear that new investigations of NC
 $\nu(\bar\nu)$--nucleon scattering are necessary in order to draw 
definite conclusions about the value of  $g_A^s$.

In this paper we calculate the contribution of the strange form
factors of the nucleon to the NC over CC neutrino--antineutrino 
asymmetry and compare our results with the information on it, which 
 one can extract from the data of the BNL--734 experiment.

In this experiment the following ratios of cross sections 
were obtained \cite{Ahrens87}:
\bea
R_\nu &=& 
\frac{\langle \sigma \rangle_{(\nu p\rightarrow \nu p)}}
{\langle \sigma \rangle_{(\nu n\rightarrow
\mu^- p)}} = 0.153 \pm 0.007 \pm 0.017
\label{rnu} \\
R_{\overline{\nu}} &=& 
\frac{\langle \sigma \rangle_{(\overline{\nu} p\rightarrow 
\overline{\nu} p)}}
{\langle \sigma \rangle_{(\overline{\nu} p\rightarrow
\mu^+ n)}} = 0.218 \pm 0.012 \pm 0.023
\label{rnubar} \\
R &=& 
\frac{\langle \sigma \rangle_{(\overline{\nu} p\rightarrow 
\overline{\nu} p)}}
{\langle \sigma \rangle_{(\nu p\rightarrow
\nu p)}} = 0.302 \pm 0.019 \pm 0.037\ ,
\label{rr}
\eea
where $\langle \sigma \rangle_{\nu(\bar\nu)}$ is a total cross section
integrated  over the incident neutrino (antineutrino) energy and weighted 
by the $\nu(\bar\nu)$ flux in a way that will be specified below. 
The first error is statistical and the second is the systematic one.

In ref.\cite{Alb96} we have shown that the measurement of
neutrino--antineutrino asymmetry 
\beq
{\cal A}_p(Q^2) = {\displaystyle
\frac{\displaystyle
\left(\frac{d\sigma}{dQ^2}\right)_{\nu p\rightarrow \nu p} -
\left(\frac{d\sigma}{dQ^2}\right)_{{\bar\nu} p\rightarrow {\bar\nu} p} }
{\displaystyle
\left(\frac{d\sigma}{dQ^2}\right)_{\nu n\rightarrow \mu^- p} -
\left(\frac{d\sigma}{dQ^2}\right)_{{\bar\nu} p\rightarrow \mu^+ n} }
}\, 
\label{asymmdif}
\eeq
will allow to obtain direct model independent information on
the axial ($F_A^s$) and magnetic ($G_M^s$) strange form factors 
of the nucleon. Indeed (\ref{asymmdif}) can be rewritten as
\beq
{\cal A}_p(Q^2) = \frac{1}{4|V_{ud}|^2}
\left(1-\frac{F_A^s}{F_A}\right)
\left(1-2\sin^2\theta_W
\frac{G_M^p}{G_M^3} -\frac{G_M^s}{2G_M^3}\right)\, ,
\label{asymff}
\eeq
where $F_A$ is the CC axial form factor and $G_M^3=(G_M^p-G_M^n)/2$
the isovector magnetic form factor of the nucleon.

Here we will use the ratios (\ref{rnu})--(\ref{rr}) 
to obtain an experimental information on the integral asymmetry. 
In the next Section we will compare this asymmetry 
with our theoretical calculation, which includes the contribution
of the strange nucleon form factors.

The Brookhaven experiment was performed using 
wide band neutrino and antineutrino beams, with an average energy
of about 1.3~GeV. Almost 
$80~\%$ of the events were due to quasielastic proton knockout
from $^{12}$C nuclei and the remaining $20~\%$ of events were due to 
elastic neutrino (antineutrino) scattering on free protons. 
In order to compare theoretical calculations with the BNL data, one must 
take into account the  energy spectrum of the neutrinos 
[$\phi_\nu(\epsilon_\nu)$] and antineutrinos 
[$\phi_{\bar\nu}(\epsilon_{\bar\nu})$]. 
In the case of elastic scattering, we define a {\it folded} 
differential cross section by:
\beq
\langle \frac{d\sigma}{dQ^2}\rangle_{\nu(\overline{\nu})p}=
\frac{1}{\Phi_{\nu(\overline{\nu})}}\int_{0.2~{\mathrm GeV}}^{5~{\mathrm GeV}}
d\epsilon_{\nu(\overline{\nu})}\left(\frac{d\sigma}{dQ^2}
\right)_{\nu(\overline{\nu})p} \phi_{\nu(\overline{\nu})}
(\epsilon_{\nu(\overline{\nu})})
\label{diffold}
\eeq
where $\Phi_{\nu(\overline{\nu})}$ is the  total neutrino
(antineutrino) flux and the limits on $\epsilon_{\nu(\bar\nu)}$ 
correspond to the experimental conditions. 

The differential cross sections are given as a function of the squared 
momentum transfer, $Q^2$, which in the case of free protons is directly 
obtained from the final proton kinetic energy in the laboratory system 
by the relation $Q^2=2MT_p$ ($M$ being the proton mass ). For 
scattering off $^{12}$C the authors of ref.\cite{Ahrens87} obtained the 
equivalent ``free scattering data''   by correcting for the Fermi motion 
and binding energy of the hit nucleon: in this case the $Q^2$ given by 
the above relation must be regarded  as an {\it effective} momentum 
transfer squared, around which the quasielastic 
$\nu(\bar\nu)$ scattering on $^{12}$C occurs.

A proper interpretation of the results in terms of scattering on the free
nucleon requires a reliable understanding of the effects associated with the 
nuclear, many--body dynamics in both the initial and final states, as well 
as with the final state interactions (FSI) between the ejected nucleon and
the residual nucleus. We have shown\cite{Alb97,Alb98} that for 
neutrino (antineutrino) energies of about 1~GeV and larger these effects
are within percentage range 
for ratios of cross sections.
Note, however, that FSI sizeably reduce ($\sim 50~\%$) the separated 
cross sections with respect to the plane wave impulse approximation 
(PWIA): indeed FSI take into account the existence of other reaction 
channels besides the quasielastic one and just approximately $50~\%$
of the reaction events correspond to elastic proton knockout.
Therefore, when applied to the individual cross sections, the interpretation 
of the BNL data as corresponding to elastic scattering on ``free'' protons
is not free from ambiguities.

In addition to the above differential cross sections, one can define 
the {\it total folded} cross sections by integrating (\ref{diffold}) 
over the (effective) momentum transfer $Q^2$:
\beq
\langle \sigma \rangle_{\nu(\overline{\nu})p} =
\int_{0.5~{\mathrm GeV}^2}^{1~{\mathrm GeV}^2} d Q^2 
\langle \frac{d\sigma}{dQ^2}
\rangle_{\nu(\overline{\nu})p} \ ,
\label{totfold}
\eeq
the limits of integration being taken from ref.\cite{Ahrens87}.

The neutrino--antineutrino {\it folded integral asymmetry}, 
$\langle {\cal A}_p \rangle$, is obtained from the neutral 
current to charge current ratio of the differences between the total 
folded neutrino and antineutrino cross sections\cite{Alb96,Alb97}:
\beq
\langle {\cal A}_p \rangle = \frac{
\langle \sigma \rangle_{(\nu p\rightarrow
\nu p)} - \langle \sigma \rangle_{
(\overline{\nu} p \rightarrow \overline{\nu} p)}}
{\langle \sigma \rangle_{(\nu n\rightarrow
\mu^- p)} - \langle \sigma \rangle_{
(\overline{\nu} p \rightarrow \mu^+ n)}}\, .
\label{asymm}
\eeq
This quantity can be written in terms of the ratios (\ref{rnu})--(\ref{rr})
as follows:
\beq
\langle {\cal A}_p \rangle =
\frac{R_\nu(1-R)}{1-RR_\nu/R_{\overline{\nu}}}
\label{asym2}
\eeq
and from the  experimental data we found
\beq
\langle {\cal A}_p \rangle = 0.136 \pm 0.008 (\mathrm{stat}) 
\pm 0.019 (\mathrm{syst})
\label{asymexp}
\eeq
where the statistical error has been estimated using the standard 
quadratic error propagation theory, while for the systematic error we 
take into account the positive correlation coefficient $\rho=0.5$ 
between systematic errors for $\nu$ and ${\bar\nu}$ cross 
sections\cite{Ahrens87}.

\section{Results and discussion}

In this Section we shall compare the experimental values for the ratios
(\ref{rnu})--(\ref{rr}) and for the asymmetry (\ref{asymexp}) with
their theoretical evaluation. We shall discuss the influence of the
strange form factors of the nucleon; moreover, for data obtained from
scattering of $\nu(\bar\nu)$ on nuclei, the influence of the nuclear 
medium will be shortly examined.

Let us first discuss the sensitivity of the integral asymmetry 
to different  assumptions. 
First of all we have considered the effect of folding the elastic 
$\nu(\bar\nu)$ cross sections with the corresponding fluxes. In 
Fig.~1 we show the integral asymmetry as a function of the 
axial strange constant, $g_A^s$. The electric and magnetic 
strange form factors have been taken to be zero. 
In the case of elastic neutrino (antineutrino)--proton scattering 
the folded integral asymmetry (solid line) is compared with the 
``unfolded''  integral asymmetry   evaluated at a fixed  
$\nu(\bar\nu)$ energy  $\epsilon_{\nu(\bar\nu)}=1$~GeV 
(empty dots). The difference between the two curves is less 
than 2~\%, a result which one could have expected from the similarity
between the neutrino and antineutrino spectra in the BNL--734 
experiment.\footnote{One could notice that the average energy of the
neutrino (antineutrino)
spectra of the BNL experiment is about 
$\epsilon_{\nu({\bar\nu})}\simeq 1.3$~GeV; however between 1 and 2 GeV
the unfolded asymmetry  varies at most by $~0.3\%$,
which makes
irrelevant the fixed value of $\epsilon_{\nu(\bar\nu)}$ that we
utilize 
for the unfolded asymmetry.}

Next we compare the results obtained for elastic scattering
on protons to the ones obtained in the impulse approximation (IA) 
for quasielastic $\nu(\bar\nu)$ scattering on $^{12}$C.
Three different approximations are 
used to describe the nuclear dynamics: first we consider a
relativistic Fermi gas (RFG) within the PWIA, namely 
without distortion of the ejected nucleon wave. For the RFG we show
in Fig.~1 both the folded (dashed line) and the unfolded (dotted line)
asymmetry. Further we use a relativistic shell model (RSM), both
within PWIA (dot--dashed line) and with inclusion of the FSI 
of the observed nucleon (three--dot--dashed line); the 
latter  is taken into account 
through a relativistic optical potential (ROP), which is employed 
together with the RSM (see refs.\cite{Alb97} and \cite{Udi96} for 
details of these  models).
Due to the small effect of the folding procedure, which can be argued
from the elastic and the RFG cases, for the RSM (without and
with the final state interaction)
only the unfolded asymmetry is shown, 
again at  $\epsilon_{\nu(\bar\nu)}=1$~GeV.
We notice that the quasielastic $\nu({\bar\nu})$--nucleus scattering,
treated within the RFG and RSM (in PWIA) 
gives results almost identical to the case of 
elastic $\nu(\bar\nu)$--proton
scattering. The effect of FSI, instead, shows up in a reduction 
of the asymmetry of at most 2~\%,
mainly due to Coulomb effects, as pointed out  in 
ref.\cite{Alb97}. 
Nevertheless the effect of  the axial strange constant $g_A^s$
on the integral asymmetry remains larger (for $-g_A^s\ge 0.05$) 
than the effects 
associated with nuclear models (including FSI) and/or with the folding 
over the $\nu(\bar\nu)$ spectra.
This result justifies previous analyses of the BNL quasielastic
data
in terms of a Fermi Gas model when ratios of cross sections
and/or asymmetries are concerned; indeed for quasielastic processes, 
ratios of cross sections are basically the same as for 
the corresponding  elastic $\nu(\bar\nu)$--proton processes. 
 However we remind that FSI are not negligible for the single
cross sections\cite{Alb97,Garv92,Garv93a}. 

On the basis of the previous discussion, in what follows we 
just consider ratios of {\it folded} 
cross sections for {\it elastic scattering} on free protons.

In Fig.~2 we demonstrate the effects of strangeness for  
the ratios (\ref{rnu})--(\ref{rr}) and the integral asymmetry 
(\ref{asymm}). The experimental values for the various
quantities are indicated by the shadowed regions: the error band 
corresponds to one standard deviation, calculated from quadratic propagation
of the statistical and systematical errors.
We have assumed the usual dipole parameterization 
both for non--strange and strange form factors, the latter being  
 $F_A^s(Q^2)=g_A^s G_D^A(Q^2)$, $G_M^s(Q^2)=\mu_s G_D^V(Q^2)$ and 
$G_E^s(Q^2)=\rho_s \tau G_D^V(Q^2)$ ($\tau=Q^2/4M^2$), where  
$G_D^{V(A)}(Q^2)=(1+Q^2/M_{V(A)}^2)^{-2}$ and we keep
 the strengths $g_A^s$, $\mu_s$ and $\rho_s$ as free parameters.
We assume the same values  for the strange cutoff masses 
as for the non--strange vector (axial) form factors.
We do not discuss here other  parameterizations for 
the $Q^2$-dependence of the strange form factors, about which
practically nothing is known (see refs.~\cite{Kolb97,Fork94}). A
decrease of $G_M^s$ and $F_A^s$ stronger than dipole at high $Q^2$ 
(as suggested by the quark counting rule) 
would obviously reduce the global effect of
strangeness. However it was shown in previous
works~\cite{Alb96,Alb97} that the effect of different
parameterizations of the strange form factors is
very small in the BNL $Q^2$ region, of the order of $\sim 1-2\%$. 

In Fig.~2(a) we show the ratios $R_\nu$ and $R_{\overline{\nu}}$ versus
$\mu_s$ for two values of the axial--strange constant: $g_A^s=0,-0.15$
and three values of the electric strange constant: $\rho_s=0,\pm 2$. 
For the axial cutoff mass we use the value $M_A=1.032$~GeV\cite{Garv93}.
Both observables are much more sensitive to the axial strange constant  
$g_A^s$, than to $\rho_s$: the former gives an effect of the order of
$\sim 15\%$ for $R_\nu$ and $\sim 27\%$ for $R_{\overline{\nu}}$. The
influence of the magnetic and electric strange form factors, instead, 
amount, respectively, to  $\sim 8\%$ ($R_\nu$), $\sim 7\%$ 
($R_{\overline{\nu}}$) for $\mu_s$ and to $\sim 4\%$ ($R_\nu$), 
$\sim 7\%$ ($R_{\overline{\nu}}$) for $\rho_s$. 
Note that the role played by the electric and magnetic
strange form factors is similar in the case of antineutrinos, whereas for
neutrinos the dependence upon $\mu_s$ is clearly  stronger.
This agrees  with the discussion presented in ref.~\cite{Alb98}. 
As it is seen from Fig.~2(a), within the present assumptions for the 
form factor parameterization and for $M_A$, in the rather
large range of $\mu_s$ and $\rho_s$  considered here, a value 
of the strange axial constant $g_A^s$ as large as $-0.15$ is not 
favoured by the BNL--734 data.

Results of the calculation of the  ratio $R$ and the integral asymmetry 
$\langle {\cal A}_p \rangle$ are
shown in Fig.~2(b): the maximum relative change in the ratio $R$, 
obtained in the ranges of strange parameters considered here,
amounts to  $\sim 10\%$ for $g_A^s$, 
$\sim 13\%$ for $\mu_s$ and $\sim 5\%$ for $\rho_s$.
Both for $R$ and $\langle {\cal A}_p \rangle$ 
the effects induced by the axial and magnetic strange form
factors are similar. 
These effects are clearly larger (in $R$) than the ones due to the
electric strange form factor. Moreover
it is worth noticing that the integral asymmetry does not depend 
at all upon the electric strange form factor, 
a result already obtained in ref.\cite{Alb96} 
for the unfolded asymmetry ${\cal A}_p(Q^2)$.
The maximum relative change in  $\langle {\cal A}_p \rangle$,
obtained in the ranges of the strange parameters considered here,
is fairly sizeable and amounts to  
$\sim 12\%$ (for $g_A^s$) and $\sim 14\%$ (for $\mu_s$). 
All the considered values of the strange parameters are compatible
with the  asymmetry  $\langle {\cal A}_p \rangle$  within the experimental
errors. However for values of $g_A^s$ as large as $-0.15$ 
the experimental value of $R$ favours $\mu_s\lesssim 0$.

The experimental relative errors for the various ratios and the asymmetry
are: $\sim 12\%$ ($R_\nu$ and $R_{\overline{\nu}}$), $\sim 14~\%$ ($R$) 
and $\sim 16~\%$ ($\langle {\cal A}_p \rangle$). 
Thus the experimental uncertainties are of the same order as the effects
of the strange form factors of the nucleon. 
At present the error bands are clearly too large to allow any definite 
conclusion on the strangeness content of the nucleon: as already pointed 
out in ref.\cite{Garv93,Hor93}, one needs to considerably reduce
the experimental errors.
 Nevertheless, the results shown in Fig.~2 give indications
that may help in the analysis and interpretation of the present 
and, in our auspices, future data.
As we have already noticed, the comparison between 
the theoretical calculations and the 
experimental data for the ratio $R_{\overline{\nu}}$ (fig.~2a) shows
that values of $-g_A^s \geq 0.15$ are clearly disfavoured: moreover
 a value $g_A^s=-0.10$ (see ref.~\cite{Ellis95})
 seems to require negative values of 
$\mu_s$, of the order of $-0.2$.
Yet the analysis of the other observables is not conclusive,
mainly because of the width of the experimental bands, which are 
compatible with many different choices of the strangeness parameters,
$g_A^s$, $\mu_s$, $\rho_s$, including the one which sets all of 
them to zero. Keeping this in mind and without any claim for 
a definitive evidence, the results seem to favour negative values 
of the magnetic strange parameter, $\mu_s$, if $-g_A^s$ is relatively
large, in agreement with the findings of ref.~\cite{Garv93}.

We remind here that a value of the strange magnetic form factor of 
the nucleon has been recently measured at BATES~\cite{Muel97}, with 
the result $G_M^s(0.1 \mathrm{GeV}^2) = 0.23 \pm 0.37 \pm 0.15 \pm 0.19$.  
This value is affected by large experimental and theoretical 
uncertainties (the last error refers to the estimate of radiative 
corrections), but it is centered around a positive $\mu_s$,
 although it is still 
compatible with zero or negative values of $\mu_s$. At present
neutrino scattering and parity violating electron scattering experiments
appear to be not conclusive for what concerns $\mu_s$.
We also notice that, if the P--odd asymmetry measured in the scattering 
of polarized electrons on nucleons 
will provide a more stringent information on the 
strange magnetic form factor, then future, precise experiments 
combining the measurement of $\nu$ and ${\bar\nu}$--proton scattering 
could allow
a determination of the axial strange form factor {\it and} of the 
electric one\cite{Alb98}. The effect of the latter is clearly 
smaller than the one associated to the axial and magnetic strange 
form factors. This is especially true for $R_{\nu}$, whereas $R_{\bar\nu}$ 
shows a non--negligible sensitivity to $\rho_s$.
 
Concerning the asymmetry [lower part of Fig.~2(b)], it is obviously 
insensitive to the electric strange form factor, according to 
its definition [see formula(\ref{asymff})]; 
it is, instead, rather sensitive to $g_A^s$ and $\mu_s$,
but, for the time being, the experimental error band does not allow us
to discriminate among the various possible choices for $g_A^s,\,\mu_s$
values.

Finally let us discuss the role played by the axial cutoff mass $M_A$.
For this purpose we show in Fig.~3(a) and 3(b)  the various observables,
$R_\nu$, $R_{\overline{\nu}}$, $R$ and $\langle {\cal A}_p \rangle$,
versus the axial strange constant $g_A^s$ for three different choices
of $M_A$: $M_A=0.996, 1.032$ and $1.068$~GeV, which are, within
$1~\sigma$, the values obtained in fit number IV of ref.\cite{Garv93}.
 The values 
of the magnetic and electric strange parameters, $\mu_s$, $\rho_s$ have 
been fixed to zero. In Fig.~3 we compare the theoretical predictions with
the ratios (\ref{rnu})--(\ref{rr}) measured in  BNL--734 experiment, 
and with the asymmetry (\ref{asymm}), 
with the same error bands shown in Fig.~2.

Solid lines correspond to results for $M_A=1.032$ GeV,
dashed lines to $M_A=1.068$ GeV and dot--dashed to $M_A=0.996$ GeV. 
One can see that by varying $M_A$ in the considered range, the
ratio  $R_{\overline{\nu}}$ is changed by $\sim 18~\%$,
 $R$ by  $\sim 10~\%$ and $R_\nu$ by $\sim 5~\%$; therefore 
these ratios rather strongly depend on the precise  value of $M_A$,
particularly in the case of antineutrinos.
 This observation was already pointed out in 
ref.~\cite{Alb98} and stands out more clearly here.
Moreover the comparison with the experimental data clearly shows, as in
ref.~\cite{Garv93}, a strong correlation between $g_A^s$ and $M_A$, which is
particularly evident in $R_{\overline{\nu}}$: keeping in mind that here 
the vector strange form factors are set to zero, the results obtained for 
$R_{\overline{\nu}}$ would disfavour, within one standard deviation, 
a wide range of values for $g_A^s$. In particular  $M_A=1.068$~GeV is 
only compatible with $g_A^s$--values 
such that $-g_A^s\lesssim 0.05$, $M_A=1.032$~GeV with 
$-g_A^s\lesssim 0.1$, whereas $M_A=0.996$~GeV extends
the range of allowed $g_A^s$--values to $-g_A^s\lesssim 0.15$. 
It is worth noticing that the other quantities shown in Fig.~3 are
much less restrictive on the  chosen values for these parameters.

In contrast with the ratios (\ref{rnu})--(\ref{rr}), we have found 
 that the neutrino--antineutrino asymmetry is practically independent
on the value of  the axial
cutoff mass $M_A$; this fact makes this quantity 
more suited to determine $g_A^s$ 
independently from the $Q^2$ behaviour of the 
axial form factors\cite{Alb96}.

In conclusion, we have thoroughly re--examined the data of the elastic
$\nu(\bar\nu)$--proton scattering BNL--734 experiment.  
We have checked that
the effects of the nuclear structure and interactions
on the ratios of cross sections and on the 
asymmetry considered here are negligible.
This is an important prerequisite to 
draw reliable conclusions on the nuclear strange 
form factors from this type of experiments.

Although our results go in the same direction as some
authors have claimed before, we can state here that
the experimental uncertainty is still too large
to be conclusive about specific values of the strange
form factors of the nucleon. A rather wide range of values for the strange 
parameters, $g_A^s$, $\mu_s$ and $\rho_s$,
is compatible with the BNL--734 data and more precise measurements are
thus needed in order to determine simultaneously the electric, magnetic
and axial strange form factors of the nucleon.
A crucial uncertainty for this determination comes from the existing
errors on the axial cutoff mass $M_A$.\footnote{Notice that from 
quasielastic $\nu(\bar\nu)$ scattering the value 
$M_A=1.09\pm 0.03\pm 0.02$\cite{Ahrens2} was found.}

Our investigation indicates that
 one can extract the strange form factors of the
nucleon  from ratios like $R_{\bar\nu}$ only if the 
axial cutoff mass will be known with much better accuracy than at
 present. On the contrary,
a high precision measurement of the neutrino--antineutrino 
asymmetry ${\cal A}_p$ could allow to determine the axial and vector
strange form factors even without the knowledge of the precise value
of $M_A$.

We can  conclude that the uncertainty 
of the available data does not allow
to set stringent limits on the strange vector and axial--vector 
parameters, but future, more precise measurements could make 
their determination possible in a model independent way.

\vspace{1cm}
\subsection*{Acknowledgments}
This work was supported in part by funds provided by DGICYT (Spain) under
contract Nos. PB95--0123 and PB95--0533--A, the Junta de Andaluc\'{\i}a,
 in part by Complutense University (Madrid) under project n. PR156/97 
and in part by funds provided by the U.S. Department of Energy (D.O.E.)
under cooperative research agreement \# DF--FC02--94ER40818.
S.M.B. acknolwledges the support of the Sonderforschungsbereich
375-95 fuer Astro-Teilchenphysik der Deutschen Forschungsgemeinschaft";
C.M. aknowledges the post--doctoral fellowship under the INFN--MIT 
``Bruno Rossi'' exchange program.



\newpage
\begin{figure}[h]
\begin{center}
\mbox{\epsfig{file=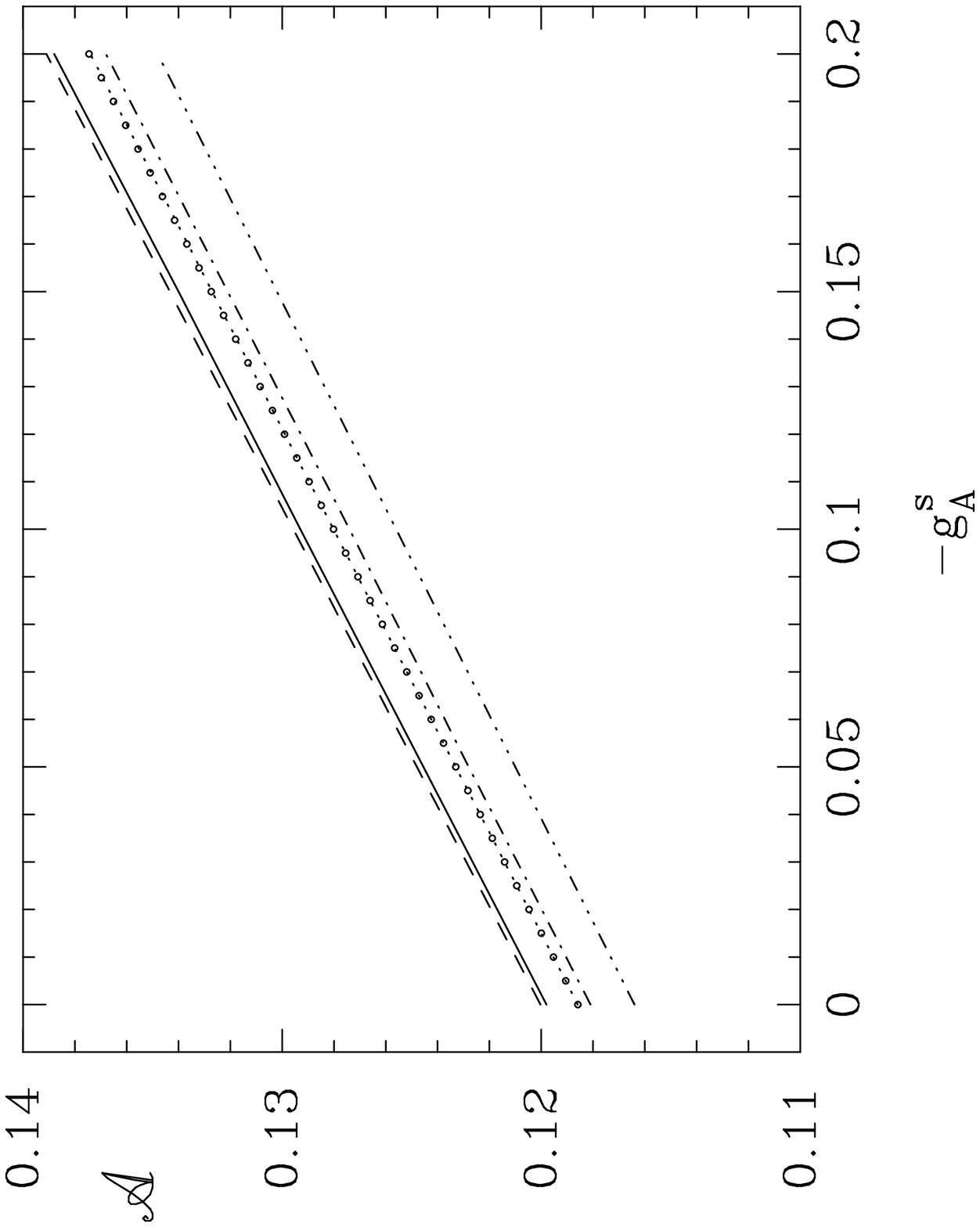,width=0.85\textwidth}}
\end{center}
\vskip -1cm
\caption[Fig.~\ref{fig1}]{\label{fig1}
The integral asymmetry ${\cal A}_p$ versus
$g_A^s$. The magnetic and electric strange form factors have
been fixed to zero. The solid line corresponds to the {\it folded} 
$\nu (\overline{\nu})$--proton 
elastic scattering asymmetry, the empty dots to 
elastic scattering without folding at
$\epsilon_{\nu (\overline{\nu})}=1$ GeV. 
Results for the quasi--elastic asymmetry on $^{12}$C  are shown by
the following curves:
dashed line (RFG, with folding), dotted line (RFG, unfolded), 
dot--dashed line (RSM, unfolded) and 
three--dot--dashed line (RSM+ROP, unfolded); all unfolded curves
are evaluated at $\epsilon_{\nu (\overline{\nu})}=1$ GeV.}
\end{figure}
    
\begin{figure}[p]
\begin{center}
\mbox{\epsfig{file=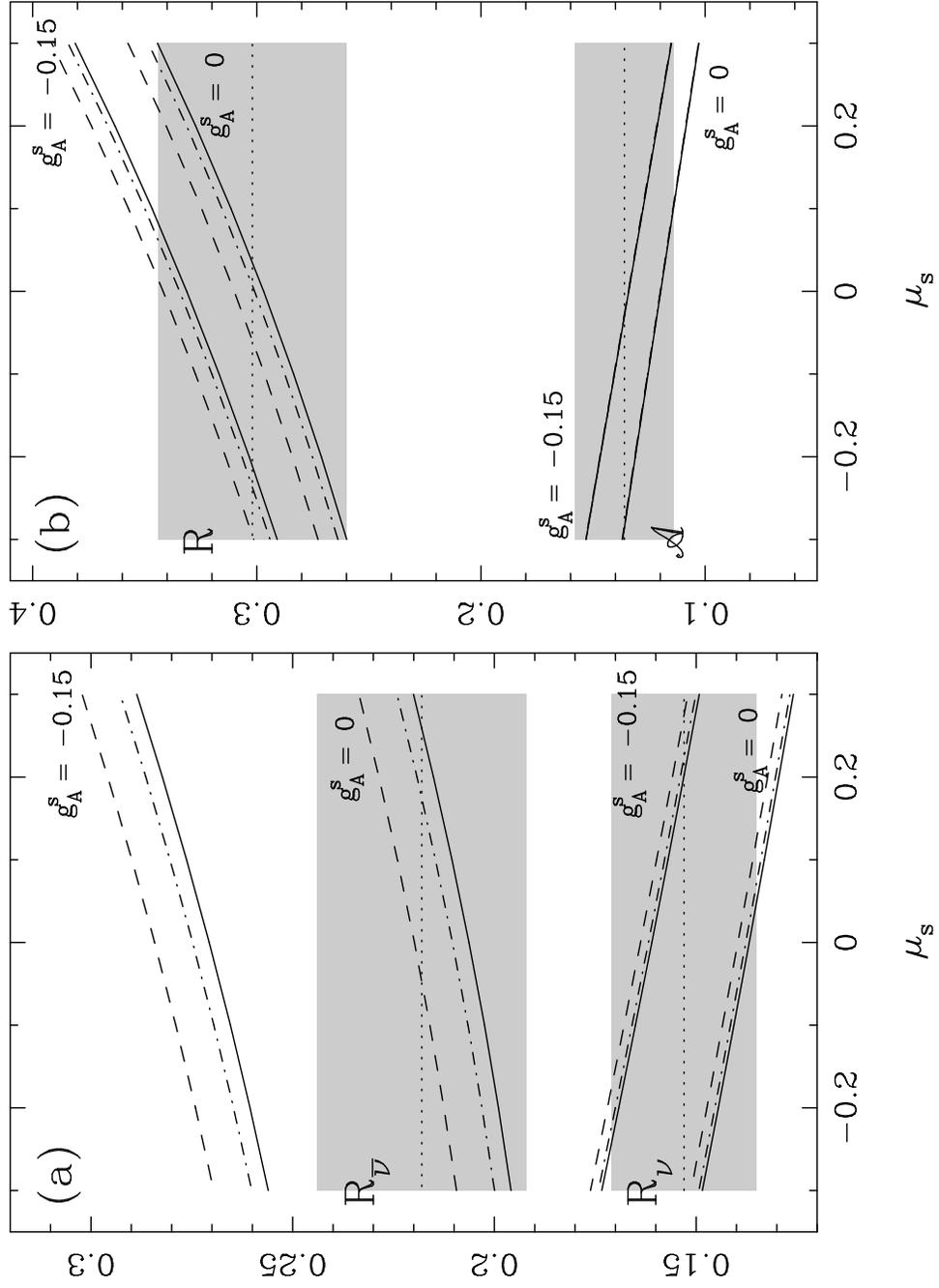,width=0.9\textwidth}}
\end{center}
\vskip -1cm
\caption[Fig.~\ref{fig2}]{\label{fig2}
The ratios $R_{\nu}$ and $R_{\bar\nu}$ (a), and
$R$ and $\langle {\cal A}_p\rangle$ (b), as a function of $\mu_s$:
all curves correspond to $\nu(\overline{\nu})$--p elastic scattering.
Results are shown for $g_A^s=0$ and $g_A^s=-0.15$. In both cases
we have chosen $\rho_s$ to be: $\rho_s=0$ (solid line),
$\rho_s=-2$ (dot--dashed line) and $\rho_s=+2$ (dashed line).
The shadowed regions correspond to the experimental data measured at
BNL-734 experiment [eqs.~(5-7,9)].}
\end{figure}

\begin{figure}[p]
\begin{center}
\mbox{\epsfig{file=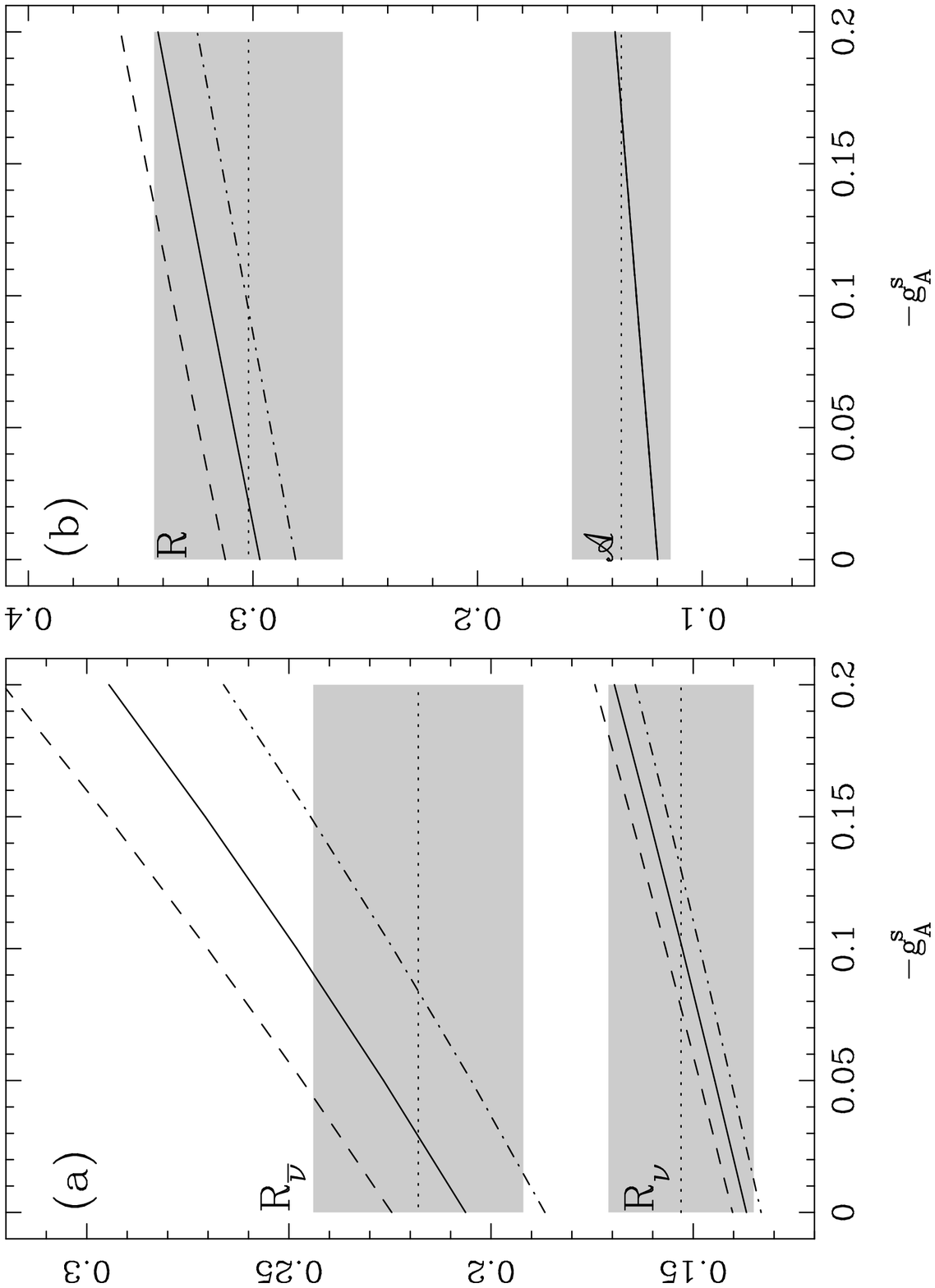,width=0.9\textwidth}}
\end{center}
\vskip -1cm
\caption[Fig.~\ref{fig3}]{\label{fig3}
The ratios $R_{\nu}$ and $R_{\bar\nu}$ (a), and
$R$ and $\langle {\cal A}_p\rangle$ (b), as a function of $g_A^s$:
all curves correspond to elastic scattering and the electric and magnetic
strange parameters have been taken $\mu_s=\rho_s=0$.
Results are shown for three different values of the axial mass cutoff:
$M_A=1.032$ GeV (solid lines), $M_A=1.068$ GeV (dashed lines) and
$M_A=0.996$ GeV (dot--dashed lines).}
\end{figure}


\begin{thebibliography}{abdce99}

\bibitem{Adams94} D. Adams {\it et. al.,}
        Phys. Lett. {\bf B329}, 399 (1994).
\bibitem{Abe95}
        K. Abe {\it et al.,} Phys. Rev. Lett. {\bf 74}, 346 (1995).
\bibitem{Ellis95} J. Ellis and M. Karliner, Phys. Lett. {\bf B341},
        397 (1995).
\bibitem{Magnon} A. Magnon, plenary talk at INPC98, Paris, August
        1998; {\it The Spin Muon Collaboration}, Phys. Rev. 
        {\bf D58}, 112002 (1998).
\bibitem{Adams97} D. Adams {\it et al.}, (SM Collab.) Phys. Rev. 
        {\bf D56}, 5330 (1997).
\bibitem{Abe95b} K. Abe {\it et al.}, (E143 Collab.) Phys. Rev. Lett.
        {\bf 75}, 25 (1995).
\bibitem{Reya98} B. Lampe and E. Reya, preprint hep-ph/9810270 (Oct. 1998).
\bibitem{Manohar} D.B. Kaplan and A. Manohar, Nucl. Phys. {\bf B310},
        527 (1988).
\bibitem{Alb96}
        W.M. Alberico, S.M. Bilenky, C. Giunti and C. Maieron,
        Z. f\"ur Physik {\bf C70}, 463 (1996).
\bibitem{PVelec} see, for example: M.J. Musolf {\it et al.}, Phys. Rep.
        {\bf 239}, 1 (1994)
\bibitem{Ahrens87} L.A. Ahrens, {\it et. al.,} Phys. Rev. 
        {\bf D35}, 785 (1987).
\bibitem{Garv93} G.T. Garvey, W.C. Louis and D.H. White, Phys. Rev.
        {\bf C48}, 761 (1993).
\bibitem{Hor93} C.J. Horowitz, H. Kim, D.P. Murdock and S. Pollock,
        Phys. Rev. {\bf C48}, 3078 (1993).
\bibitem{Barb96} M.B. Barbaro, A. De Pace, T.W. Donnelly, A. Molinari
        and M.J. Musolf, Phys.  Rev. {\bf C54}, 1954 (1996).
\bibitem{Alb97}
        W.M. Alberico, M.B. Barbaro, S.M. Bilenky, J.A. Caballero, C. Giunti, 
        C. Maieron, E. Moya de Guerra and J.M.Udias, Nucl. Phys. {\bf A623},
        471 (1997).
\bibitem{Alb98}
        W.M. Alberico, M.B. Barbaro, S.M. Bilenky, J.A. Caballero, C. Giunti, 
        C. Maieron, E. Moya de Guerra and J.M.Udias, Phys. Lett. {\bf B438},
        9 (1998).
\bibitem{Udi96} J.M. Ud\'{\i}as, P. Sarriguren, E. Moya de Guerra and J.A.
        Caballero, Phys. Rev. {\bf C53}, R1488 (1996); ibid. {\bf C48},
        2731 (1993).
\bibitem{Garv92} G.T. Garvey, S. Krewald, E. Kolbe and K. Langanke, 
        Phys. Lett. {\bf B289}, 249 (1992).
\bibitem{Garv93a} G.T. Garvey, E. Kolbe, K. Langanke and S. Krewald,
        Phys. Rev. {\bf C48}, 1919 (1993).
\bibitem{Kolb97} E.Kolbe, S.Krewald and H.Weigel, 
        Z. f\"ur Physik {\bf A358}, 445 (1997).
\bibitem{Fork94}
        H. Forkel et al., Phys. Rev. C {\bf 50}, 3108 (1994);
        H.C. Kim, T. Watabe and K. Goeke, Nucl. Phys. A 
        {\bf 616}, 606 (1997); M. Kirchbach and D. Arenh\"ovel, 
        Proceedings,
        {\it Physics with GeV--particle beams}, 414 (1994)
        (hep--ph/9409293).
\bibitem{Muel97} B. Mueller {\it et al.}, SAMPLE Collaboration,
        Phys. Rev. Lett. {\bf 78}, 3824 (1997).
\bibitem{Ahrens2} L.A. Ahrens {\it et al.}, Phys. Lett. {\bf B202},
        284 (1988).
\end{thebibliography}
\end{document}